# SIGNING INITIATIVE PETITIONS ONLINE: POSSIBILITIES, PROBLEMS, AND PROSPECTS

**SUMMARY**

Many people expect the Internet to change American politics – most likely in the direction of increasing direct citizen participation and forcing government officials to respond more quickly to voters' concerns. An initiative petition with these objectives is currently circulating in California that would authorize use of electronic signatures over the Internet to qualify candidates, initiatives, and other ballot measures.

Proponents of Internet signature gathering say it will significantly lower the cost of qualifying initiatives and thereby reduce the influence of organized, well-financed interest groups. They also believe it will increase both public participation in the political process and public understanding about specific measures. However, opponents question whether Internet security is adequate to prevent widespread abuse and argue that the measure would create disadvantages for those who lack access to the Internet. Some observers also express concern that Internet petition signing would make qualifying ballot measures too easy and thus further distance the initiative process from the deliberative political discourse envisioned by the framers of the U.S. and California constitutions.

Petition signing on the Internet would draw on the technologies and processes developed for electronic commerce ("e-commerce") and "e-government" transactions. Recent federal and California legislation authorizes the use of computer-generated electronic signatures for such transactions. Registered voters would be able to view and download information about proposed initiatives on Internet websites, sign petitions on a computer, and transmit their signatures over the Internet to be counted toward the total needed for ballot qualification.  For greater security, Internet petition signing would likely use "digital signatures" that employ advanced encryption techniques, possibly on "smart cards" containing computer chips, as well as state-approved "certification authorities" that vouch for the signer's identity.  Digital signatures would then be decrypted and matched against the current list of registered voters under the supervision of state and county election officials.

However, security problems of networked computers make Internet petition signing potentially vulnerable to fraud and other abuse.  Digital signatures protected only by passwords may be easily lost, stolen, copied, or otherwise compromised.  Similar vulnerabilities prompted the California Internet Voting Task Force in January 2000 to recommend against the early implementation of Internet voting from remote computers.  Internet petition signing may involve less risk than Internet voting, largely because a signer's anonymity need not be preserved and officials could recheck a sample of signatures by contacting voters directly.  Moreover, the automated decryption and checking process for digital signatures may prove superior to today's manual methods for verifying handwritten signatures.

Although few e-commerce and e-government transactions in the United States today use digital signatures, smart cards, or certification authorities, the industry is investing heavily to enhance online security.  Given the commercial pressure to reduce risks and losses from large numbers of Internet transactions, identification and security methods will undoubtedly improve, and it seems highly likely that the commercial world will find workable solutions.  Additional efforts to develop Internet voting, both for government and non-government elections, will also spur the development of better security approaches. Even so, the security standards must be tighter for Internet voting or petition signing than for e-commerce in order to maintain public trust in the election process.

The costs associated with Internet petition signing include those required to issue smart cards, digital signature certificates, and encryption keys to voters; the costs of developing the infrastructure to renew certificates and to revoke and reissue them if compromised; and the costs of developing Internet-accessible voting lists for signature verification. The initial infrastructure costs for 25 million California adults would be upwards of $200 million, with additional recurring costs of managing the system. Once this infrastructure is in place, however, the costs of gathering and processing petition signatures should decrease, perhaps significantly. Changes in absentee voting and other procedures prompted by the November 2000 election problems may also provide part of the infrastructure needed for online petition signing. The issues surrounding voting are closely intertwined with those for petition signing; consequently, future studies of Internet voting should also consider the implications for Internet petition signing.

Although limited access to the Internet remains a problem, its magnitude seems to be diminishing over time. Survey data report that more than two-thirds of California adults were Internet users as of October 2000 but that Internet use varies considerably by race and ethnicity, income, education level, age, and region. Although these gaps are steadily shrinking, market and demographic factors alone will not bring all Californians online. As a consequence, any near-term implementation of Internet petition signing should include access provisions for those who are not connected to the Internet at home, school, or work.

Beyond the issues of security, cost, and access lie larger questions about the effects of Internet signature gathering on direct democracy. Would it encourage greater and more informed public participation in the political process?  Or would it flood voters with ballot measures and generally worsen current problems with the initiative process itself?  Because we lack good data on these questions and systematic studies of them, today's answers to them are largely conjectural.  We simply do not understand the full implications of using the Internet for petition signing or voting.   We can be fairly sure, however, that Internet petition signing, like Internet voting, will have unintended consequences.  That may be reason enough for many to oppose its early implementation in California, but it will not make the concept disappear.  Its proponents are likely to gain strength as young people who have grown up with the Internet reach voting age.  Internet petition signing seems to be an idea whose time is not yet ripe but is clearly ripening.  Its emergence on the political horizon should spur reformers of the initiative process to get on with their work before they are overtaken by events in cyberspace.

# 1. INTRODUCTION[1]

"The Internet changes everything" is a mantra familiar to technologists, entrepreneurs, and the media.[1] Indeed, the Internet has already transformed many organizations and business sectors and profoundly affected others. These trends suggest to many that the Internet will inevitably change American politics – most likely in the direction of increasing direct citizen participation and forcing government officials to respond more quickly to voters' concerns. Certainly the dramatic vote counting problems in the 2000 presidential election have brought new calls for using the Internet in state and federal elections.[2] Although attention has focused primarily on Internet voting, efforts are also under way to authorize the use of electronic signatures over the Internet to qualify candidates, initiatives, and other ballot measures. An initiative petition is currently circulating in California that would submit such a plan to voters in March 2002.[3]

Petition signing on the Internet would draw on the technologies and processes developed for electronic commerce ("e-commerce"). It would also draw on the growing use of the Internet for disseminating government information and facilitating online communications and transactions between citizens and government ("e-government"). Its proponents claim that Internet signature gathering will significantly lower the cost of qualifying initiatives and thereby reduce the influence of organized, well-financed interest groups.  They also maintain that Internet petition signing will increase both public participation in the political process and public understanding about specific measures.  However, questions about security and access pose significant problems for Internet signature gathering, as they do for casting and counting ballots using the Internet.[4]  Some observers also express concern that Internet petition signing would make qualifying initiatives *too* easy and thus further distance the initiative process from the deliberative political discourse envisioned by the framers of the U.S. and California constitutions.

---

[1] A previous version of this paper was presented to the Speaker's Commission on the California Initiative Process 1n January 2001. The author has benefited from helpful comments from Robert Anderson, Mark Baldassare, Max Neiman, Joyce Peterson, Fred Silva, Willis Ware, and Jeri Weiss. I also thank the Public Policy Institute of California for its support.

[1] An early example of this now-popular phrase comes from Cortese, Amy, "The Software Revolution: The Internet Changes Everything," *Business Week*, December 4, 1995 http://www.businessweek.com/1995/49/b34531.htm.

[2] For example, see Cooper, Audrey, "Legislator Proposes Online Voting for California," Associated Press, December 5, 2000; and Chambers, John, "Can Technology Fix Balloting Problems? Yes; Harness Strength of the Internet," *USA Today*, December 19, 2000.

[3] "Digital Signature. Election Petitions. Public and Private Transactions. Initiative Statute." Summary available at http://www.ss.ca.gov/elections/elections_j.htm#2000General; the full text can be found on the "Smart Initiatives" website <http://www.smartinitiatives.org/English/civiset.html>.

This paper explores the prospects for and issues surrounding Internet petition signing in California.  After describing how voters would use the Internet to "sign" petitions and how their electronic or digital signatures could be verified, it goes on to discuss security, cost, access, and equity issues that pose significant obstacles to online petition signing.  It then outlines trends in Internet voting, e-commerce, and e-government that may affect the development of Internet petition signing.  The final section discusses some broader implications of the Internet for the initiative process, summarizes the arguments pro and con, and concludes that while Internet petition signing is not ready to be implemented in the next election cycle, public pressure to authorize it will continue to build and could prove unstoppable over the next few years.

---

[4] The Final Report of the California Internet Voting Task Force, convened by Secretary of State Bill Jones, provides a detailed discussion of security and related issues.  See Secretary of State, State of California, *Final Report of the California Internet Voting Task Force, Appendix A*, January 18, 2000; available at http://www.ss.ca.gov/executive/ivote/.

## 2. HOW WOULD INTERNET PETITION SIGNING WORK?

Initiative petitions must receive a specified number of valid signatures from registered voters to be placed on the ballot.[5] Proposals of Internet petition signing would change existing election laws to permit registered voters to sign petitions on a computer and transmit their signatures over the Internet to be counted toward the required total. Nearly all such proposals would permit signing at any computer, so long as proper security procedures were followed. At least for the foreseeable future, however, Internet petition signing would complement rather than supplant conventional methods of gathering written signatures.

Internet signature gathering requires at least the following three technical components:
- One or more websites that display the text of the proposed initiative on the public Internet;
- Means for voters to sign the initiative petition and transmit their signatures to the officials certifying them; and
- Means to authenticate the signatures and check them against the lists of registered voters.

Such websites could be run either by the initiative proponent or by state election officials. Under current California law, no changes in a proposed initiative are permitted once it has been approved by the Attorney General's office for signature gathering. Accordingly, such websites should display the initiative in a format that is widely accessible but not readily alterable, such as Adobe Acrobat®. Of course, these websites must be secured against hacker intrusion, denial of service attacks,[6] and other abuses;[7] but these problems appear to be less critical than those of securely gathering and authenticating voters' signatures on the Internet.

---

[5] California requires signatures equivalent to 5 percent of the vote in the most recent gubernatorial election for statutory initiatives (419,260) and 8 percent for constitutional initiatives (670,816).

**Electronic and Digital Signatures**

Internet petition signing would build on the acceptance of electronic signatures for contracts and many other transactions as authorized under the 1999 California Uniform Electronic Transactions Act[8] (UETA) and the federal Electronic Signatures in Global and National Commerce (E-SIGN) Act of 2000.[9] These laws basically state that a signature, document, or record may not be denied legal effect or enforceability solely because it is in electronic form.[10]  The laws deliberately do not specify the methods to be used for electronic signatures or the level of security required.

California's UETA statute broadly defines an electronic signature as "an electronic sound, symbol or process attached to or logically associated with an electronic record and executed or adopted by a person with the intent to sign the electronic record."[11] Thus, a customer can make a legally binding purchase simply by clicking on an icon shown on the computer screen so long as the parties have agreed to conduct the transaction using electronic media.[12] This kind of arrangement underlies much of the consumer commerce conducted on the Internet.

---

[6] Although denial of service attacks are very real threats to election websites, they pose a more serious problem to Internet voting, which is conducted over a short period of time, than to initiative signature gathering, which is carried out over several months.

[7] For example, hackers may be able to divert traffic from a legitimate website to one with a similar look that they have created; they could then fool users into revealing passwords, credit card numbers, or other personal information.

[8] *Uniform Electronic Transactions Act: California Civil Code*, California Senate Bill 820, Enacted September 16, 1999.

[9] *Electronic Signatures in Global and National Commerce Act*, P.L.106-229, Enacted June 30, 2000.

[10] Wills, testamentary trusts, and certain other specified transactions are excluded under UETA and E-SIGN.

[11] *Uniform Electronic Transactions Act,* op. cit.

The term "digital signature," although often used as a synonym for "electronic signature," more precisely denotes a technical approach for binding an electronic signature to a particular electronic record that includes protections against alteration or other abuse.[13] Digital signatures use a mathematically robust method of encryption, known as "public key cryptography," associated with a "public key infrastructure" (PKI), to ensure the integrity of electronic signatures and records transmitted over the Internet.[14] Thus, for security reasons, many proponents of Internet voting and petition signing, including the backers of the California "Smart Initiatives Initiative" now being circulated, would require the use of PKI digital signatures.

To use PKI digital signatures for petition signing, registered voters would be assigned a unique pair of private and public cryptographic keys by a public agency such as the Department of Motor Vehicles (DMV) or an approved private "certification authority."[15] The private key would be downloaded onto the voter's computer or stored on a "smart card" containing a microchip, while the public key would be registered with the certification authority. A voter could then use his or her private key to sign a petition – either on the voter's computer or on another computer or device with a smart card reader -- and send the encrypted signature[16] to the initiative website. Signatures would be decrypted using the public key registered with the certification authority and verified by election officials against the current voter list.

---

[12] E-SIGN, §101(c), states that the parties must have "affirmatively consented" to the electronic transaction.
[13] Information Security Committee, "Digital Signature Guidelines," American Bar Association, Section of Science and Technology, Electronic Commerce and Information Technology Division, 1996, http://www.abanet.org/scitech/ec/isc/digital_signature.html.
[14] Computer Science and Telecommunications Board, *The Internet's Coming of Age*, Washington, DC: National Academy Press, 2000, pp. 5/15-19; and ________, *Trust in Cyberspace*, Washington, DC: National Academy Press, 1999, pp. 124-132. California regulations approved in 1998 for digital signatures valid for use by public entities also permit the use of a technical method known as "Signature Dynamics," which requires special hardware and expert handwriting analysis. Because Signature Dynamics is more cumbersome and expensive and less secure than PKI, this discussion assumes that the PKI approach would be used for petition signing with digital signatures. See Secretary of State, State of California, "California Digital Signature Regulations," June 12, 1998, www.ss.ca.gov/digsig/regulations.htm.
[15] The list of private certification authorities approved by the California Secretary of State can be found at www.ss.ca.gov/digsig/cert1.htm.

Despite their mathematical complexity, PKI digital signatures are now used in some e-commerce and e-government transactions with relatively little added burden to either party. Private firms have established themselves as certification authorities, and several have been approved by the California Secretary of State for use by public agencies. The PKI digital signature approach to Internet petition signing thus appears technically feasible, although it raises a number of security, cost, and access issues which are discussed in the next section.

---

[16] Technically, the "signature" is the result of a mathematical calculation using the bits contained in the private key and the electronic record (petition).

# 3. SECURITY, COST, AND ACCESS ISSUES

## Security Issues Surrounding Internet Petition Signing

Newspapers regularly report the exploits of hackers who have broken in to supposedly secure computer networks, reminding us that perfect security will never be achieved in computer systems or any other human endeavor.[17] Internet petition signing is potentially vulnerable at several points and levels of the process. Websites displaying initiatives can be altered, "spoofed," or made unreachable for extensive periods of time. Private keys are usually protected by passwords that may be all-too-easily accessible or otherwise compromised. Thus, a voter's private key can be willingly or unwittingly given to someone else or copied remotely by a sophisticated intruder, who can then use it to sign petitions.[18] Viruses or other malicious code can be introduced to copy a private key or substitute another. Smart card readers can be similarly compromised. Individuals working for a certification authority, or election officials can be corrupted. The list of possible security breaches goes on.

These vulnerabilities are similar to those identified in numerous prior reports and discussions about Internet voting, such as the January 2000 final report of the California Internet Voting Task Force and the report from a National Science Foundation sponsored workshop on Internet voting held in October 2000.[19] The

---

[17] For a sensible and accessible introduction to computer security, see Culp, Scott, "The Ten Immutable Laws of Security," October 2000, www.microsoft.com/technet/security/10imlaws.asp.

[18] As computer security experts have pointed out, digital signatures can only verify that a private key assigned to an individual was securely linked by a computer to a particular electronic document or record. It does not prove that the individual intended to sign the document, or that he or she was even present when the document was signed. See Schneier, Bruce, "Why Digital Signatures Are Not Signatures," *CRYPTO-GRAM*, November 15, 2000 http://www.counterpane.com/crypto-gram-0011.html.

[19] *Final Report of the California Internet Voting Task Force*, op. cit.; Internet Policy Institute, "Report of the National Workshop on Internet Voting," March 6, 2001, http://www.internetpolicy.org/research/e_voting_report.pdf.

California Task Force concluded that "technological threats to the security, integrity and secrecy of Internet ballots are significant" and recommended against early implementation of remote Internet voting from home and office computers. Although the Task Force "did not consider Internet petition signing at any great length," its Technical Committee was concerned about the possibility of large-scale, computerized, "automated fraud" if individuals could register to vote remotely over the Internet without appearing personally and showing some sort of identification.[20] Regarding Internet petition signing, the Technical Committee report commented:

> Systems that would allow online petition signing from a home or office PC are vulnerable to malicious code or remote control attacks on the PC that might prevent the signing of a petition, or spy on the process, or permit additional petitions to be signed that the voter did not intend to sign, all without detection. Hence, for the same reasons that we do not recommend Internet voting from machines not controlled by election officials, we cannot recommend similar systems for petition-signing until such time as there is a practical solution to the general malicious code problem and the development of a system to electronically verify identity.
>
> While there are similarities between voting and petition signing, it is important to note that the two are not identical and they have somewhat different cost and security properties:
>
> - Petition signing is a year-round activity, whereas voting occurs during a limited time window. Hence, servers and other infrastructure needed to support petition signing would need to be running year-round, instead of just during a time window before election day. This may dramatically increase the total cost of managing the system.

---

See also Shamos, Michael Ian, "Electronic Voting: Evaluating the Threat," 1993, http://www.cpsr.org/conferences/cfp93/shamos.html; Mercuri, Rebecca, "Electronic Voting", http://www.notablesoftware.com/evote.html; and Rubin, Avi, "Security Considerations for Remote Electronic Voting over the Internet," 2000, http://avirubin.com/e-voting.security.html.
[20] *Ibid.*, Appendix A, pp. 9-12.

- While it is reasonable to expect voters, for security reasons, to submit a signed request for Internet voting authorization each time before they vote (similar to a request for an absentee ballot), it is not reasonable to expect voters to submit such a request each time they wish to sign a petition. As a result, voters who wish to sign petitions electronically would likely have to be issued authorization (means of authentication) that is open-ended in time. The longer such authorizations are valid, the more likely it is that some of them will be compromised, or sold, reducing the integrity of the petition-signing system over time.

- Voters can sign any number of petitions in an election cycle. Hence, a compromised authorization to sign petitions would be usable for signing any number of petitions, magnifying the damage to the system's integrity.[21]

Although these three bulleted objections should not be minimized, e-commerce sites face similar problems and are successfully using encrypted electronic signatures to deal with them (see Section 4). Of course, e-commerce firms can apply risk management concepts and tools to keep losses from security lapses at an acceptable level, whereas public trust in the initiative process may well require a higher standard. The questions then become: How secure must Internet petition signing be to gain voters' trust, and can that level of security be achieved at acceptable cost?

The security implications of Internet petition signing are not entirely negative. Compared to present methods, it could also improve verification of voter signatures. In California, county clerks now examine a random sample of 500 signatures or 3 percent of the total, whichever is higher. The results by county are given to the Secretary of State, who uses them to project a statewide total of valid signatures.[22] With Internet petition signing, every digital signature, not just

---

[21] *Ibid*, Appendix A, pp. 13-14.
[22] "The Secretary of State projects the rate [of signatures qualifying] for each county, totals the projected valid signatures from all 58 counties, and qualifies the initiative if there are 110 percent or more of the needed signatures. If the total falls between 95 and 110 percent, each signature must be individually

a sample, can be checked when decrypted to verify that the signer is a registered voter and has not previously signed the petition.[23] Consequently, statewide results should be more accurate and available more quickly. For added security, an automated query might be sent to a sample of electronic signers at their registered postal or e-mail addresses, asking them to confirm by return mail or e-mail that they actually had signed the petition.

**The Costs of Internet Petition Signing**

Advocates of Internet petition signing forecast dramatically lower costs both for initiative proponents and for county and state offices that process their petitions. Using paid signature gatherers, proponents now typically spend more than $1 million to qualify a statewide initiative in California.[24] According to Marc Strassman, Executive Director of the Smart Initiatives Project, that expense could fall "to the ten thousand dollars needed to build a first-class website, thereby allowing individuals and groups without million dollar budgets to participate in the initiative process."[25] However, initiative proponents would still incur the costs of circulating other petitions for handwritten signatures and of managing the campaigns of initiatives that qualified for the ballot.[26] Nevertheless, significant cost savings are plausible once the infrastructure for Internet petition signing is in place.

---

verified; below 95 percent, the initiative does not qualify." Simmons, Charlene, "California's Initiative Process: A Primer," California Research Bureau, California State Library, CRB-97-006, May1997, p. 10.
[23] This verification process assumes that counties and the state maintain up-to-date computer voting lists, and that the digital signatures have not themselves been compromised.
[24] Simmons, op. cit., p. 9.
[25] Strassman, Marc, "After Florida, What?" Smart Initiatives Online Newsletter, November 12, 2000.
[26] The Internet can also serve as a fundraising and organizing tool for initiative proponents (and opponents) once a measure has been qualified. Both political candidates and organized interest groups are already making effective use of the Internet for these purposes.

How much would the infrastructure cost, and who would pay for it? Marc Strassman estimates that the initial cost to the state of providing smart cards and digital certificates for roughly 25 million California adults would be less than $200 million, or about $8 per person.[27] This figure does not include the cost of smart card readers, which are widely available in cell phones and point-of-sale terminals in Europe and parts of Asia but not yet in the United States. The U.S. lag results in large part from our pervasive use of credit cards that are routinely and inexpensively checked over the telephone network for each transaction. This practice has so far obviated the need for more costly smart cards.

A smart card reader costs between $40 and $80 if bought as a separate unit but only $10 to $20 each if purchased in large quantities and integrated into cell phones or personal computers (PCs).[28] Hardly any PCs sold in the U.S. now come equipped with smart card readers, however, and PC manufacturers are unlikely to include them as standard features in the next few years. Using a cell phone or other mobile device equipped with a smart card reader to access the Internet is a likely scenario for consumer transactions; but this scenario is rather less likely for petition signing. As a consequence, ensuring general public access to smart card readers might require the state to purchase thousands of card readers, which it would then connect to the Internet at public kiosks, libraries, government offices, and other places where petitions could be signed.

Once a PKI infrastructure is in place, there will be continuing costs to manage the certification process for digital signatures. Certificates should be renewed on a regular basis to deter the potential fraud problems identified by the California Internet Voting Task Force. If an individual's private key is lost or compromised, it must be revoked and a new key pair and certificate issued. Moreover, the list of revoked certificates must be distributed promptly to election officials and anyone else who might rely on their authenticity. These recurring costs are difficult to estimate today because no PKI system of the proposed size is operational. The costs could be significantly lower if the key pairs and certificates issued for petition signing were also used for other public or private transactions, but this arrangement would further increase the risks of compromise and fraud.[29]

---

[27] Strassman, Marc, "Fuzzy Math for Smart Initiatives," Smart Initiatives Online Newsletter, December 14, 2000.
[28] Davis, Donald, "Where There's A Web, There's A Way," *CardTechnology.com*, October 2000 http://www.cardtech.faulknergray.com/news.htm.

Developing secure, up-to-date, and Internet-accessible voting lists for checking and verifying digital signatures represents another cost to state and county government. Satisfying all three criteria is not a trivial task and would likely involve substantial expense. However, it is not wholly unprecedented; Michigan has recently built an integrated statewide computer system for cross-checking voter records.[30] California's voting lists also appear to be in better shape than those of many other states. Once Internet-accessible voter lists were available and election officials were trained to use them, the cost of verifying signatures should drop appreciably below that for the existing labor-intensive method.[31]

**Access and Equity Issues**

A persistent objection to Internet petition signing is that it would create further disadvantages for the poor, minorities, and people with disabilities who do not have easy access to computers and the Internet. If online signature gathering makes it cheaper and easier to qualify initiatives, the argument goes, it will favor the wealthy, highly educated, and mostly white voters who already have Internet connections at home and work.

---

[29] Computer Science and Telecommunications Board, *Trust in Cyberspace, op. cit.*, p. 131.
[30] Harwood, John, "Fixing the Electoral System: Lessons From States Hold Hope for Reform," *The Wall Street Journal*, December 22, 2000.
[31] Strassman, "Fuzzy Math…," op. cit., reports cost estimates from California county officials of 60 cents to one dollar per signature for manual checking and verification.

Overall, Californians rank well above the national averages in terms of computer and Internet use. Surveys conducted by the Public Policy Institute of California (PPIC) indicate that as of October 2000, 68 percent of California adults were using the Internet compared with 60 percent of all U.S. adults.[32] More than half (51 percent) of the adults surveyed reported that they went online "often," a substantial increase from 43 percent in December 1999.

Even so, the most recent national[33] and California[34] data show substantial differences in computer ownership and Internet use according to race or ethnicity, income, education level, age, and region. Among California adults, differences of more than 10 percent in Internet use separate Blacks and Latinos from Asians and non-Hispanic whites (Table 1). And to no one's surprise, Internet use is characterized by a large generation gap: Californians between the ages of 18 and 64 are two and a half times more likely to use the Internet than those over 65.

In many respects, however, the "digital divide" has narrowed appreciably in the past two years. According to national data, the gender gap among Internet users has essentially disappeared.[35] In California, the gap between Latinos and non-Hispanic whites who have been to college has nearly closed, although it remains for those without some college education.[36] The generation gap is also shrinking steadily, but it will probably take two to four years before more than half of Californians age 65 and over are Internet users.[37] Given these remaining disparities, any near-term implementation of Internet petition signing should include access provisions for those who are not connected to the Internet.

---

[32] The results from seven statewide surveys of California adults from September 1999 to October 2000 are available on the PPIC website http://www.ppic.org.
[33] National Telecommunications and Information Administration (NTIA), *Falling Through the Net: Toward Digital Inclusion*, U.S. Department of Commerce, 2000, http://search.ntia.doc.gov/pdf/fttn00.pdf.
[34] Public Policy Institute of California (PPIC), "California's Digital Divide," November 2000, http://www.ppic.org/facts/digital.nov00.pdf; "PPIC Statewide Survey: Californians and Their Government--October 2000," pp. 27-28, http://www.ppic.org/publications/CalSurvey15/survey15.pdf.

**Table 1. Percentage of California Adults Using Computers and the Internet**

| Category | Computer Users | Internet Users |
|---|---|---|
| All California adults | 78* % | 68* % |
| Race/Ethnicity | | |
|   Non-Hispanic White | 80* | 71* |
|   Asian | 91 | 82 |
|   Black | 76 | 60 |
|   Latino | 71* | 56* |
| Income | | |
|   Under $20,000 | 48 | 33 |
|   $20,000 – 59,000 | 76 | 62 |
|   $60,000 and above | 93 | 85 |
| Education | | |
|   High school or less | 56 | 39 |
|   Some college | 81 | 68 |
|   College graduate | 89 | 81 |

---

[35] Results from a national survey conducted in August 2000 showed only a 0.2 percent difference between men and women using the Internet. National Telecommunications and Information Administration, *op. cit.*, p. xvi.

[36] Public Policy Institute of California, "California's Digital Divide," op. cit. The NTIA study finds that, among Blacks and Hispanic households at the national level, lower income and education appear to account for about two thirds of the reported gaps. National Telecommunications and Information Administration, *op. cit.*, pp. 14-15.

[37] According to the NTIA national data, in the 20 months between December 1998 and August 2000, Internet use among those aged 62 to 65 increased by more than 60 percent. *Ibid.*, Figure II-2, p. 36. Applying this growth rate to Californians aged 62 to 65, whose participation is already greater than 28 percent, suggests that a majority of Californians aged 65 and over will be Internet users within three years.

| | | |
|---|---|---|
| Age | | |
|   18-64 | 83 | 70 |
|   65+ | 39 | 28 |
| Region | | |
|   San Francisco Bay Area | 82 | 72 |
|   Los Angeles County | 74 | 59 |
|   Southern California | 77 | 66 |
|   Central Valley | 72 | 61 |

\* PPIC survey data from October 2000. All other figures are averages from seven PPIC surveys between September 1999 and October 2000.

# 4. IS INTERNET PETITION SIGNING INEVITABLE? TRENDS IN INTERNET VOTING, E-COMMERCE, AND E-GOVERNMENT

Proponents of Internet signature gathering argue that the Internet is an unstoppable force that is transforming all private and public sector activities and will soon be used for petition signing, voting, and other political processes. Because this outcome is inevitable, they contend, citizens and government officials should start planning to integrate Internet petition signing into the political system in ways that will best support core democratic values. This section discusses trends and developments in Internet voting, e-commerce, and e-government and the extent to which they may spur public interest in and acceptance of Internet petition signing.

**Internet Voting in Government Elections**

Internet voting in U.S. elections dates back to 1997, when astronaut David Wolf had his ballot e-mailed from his local election district in Texas to the Russian space station Mir, where he was temporarily assigned.[38] Three years later, few Internet votes were officially counted in the 2000 elections, but the topic is receiving considerable attention in the press and among citizen groups and public officials.

In a pilot project conducted by the Department of Defense, some 84 overseas military service personnel cast absentee ballots over the Internet in the 2000 presidential election. Using PKI encryption over secure circuits developed for military communications, the ballots were sent electronically to voting officials in four states – Florida, South Carolina, Texas and Utah – and were counted along

---

[38] Counting Wolf's vote required passage of special legislation by the Texas legislature. See "Hurtling Toward Cyber-Elections," Voting Integrity Project, 1999,

with other absentee ballots. Although criticized in Congress for its high cost, the pilot project "maintained the integrity if the electoral process, and in many respects posed fewer risks to election integrity than the current [overseas] absentee by-mail process," according to a DOD sponsored assessment.[39] The project director called it "a resounding success."

A significantly larger test took place in the March 2000 Arizona Democratic primary, in which nearly 40,000, or 46 percent, of the 86,000 votes were cast over the Internet.[40] Registered Democrats received a unique Personal Identification Number (PIN) in the mail and could vote from computers at 124 public polling places as well as from their homes or offices. Internet voters entered their PINs along with their names and addresses when they logged onto the primary website, and the information was checked against the voter registration list and assigned PINs. Digital signatures and other encryption techniques were not used. The binding primary election was administered by election.com, a for-profit firm specializing in Internet voting. Some technical problems arose during the four-day period for Internet voting;[41] but according to the company, no significant security breaches occurred. Voter participation was substantially higher than that for the 1996 Presidential primary, and the Arizona Democratic Party seems quite satisfied with the results. Others, however, have criticized the Arizona Democratic primary for its lack of strong security measures and election official oversight of those who voted online from remote computers.[42]

---

http://www.voting-integrity.org/projects/votingtechnology/internetvoting/ivp_3_hurtling.htm.
[39] Department of Defense, "Voting Over the Internet: Pilot Project Assessment Report," June 2001, p. 4-2, http://www.fvap.ncr.gov/voireport.pdf.
[40] election.com, "Arizonans Register Overwhelming Support for Online Voting," March 12, 2000, http://votation.com/us/pressroom/pr2000/0312.htm. Besides the 46 percent who used the Internet, 32 percent voted by mail and 24 percent went in person to the polls.
[41] Jesdanun, Anick, "Resistance Continues for Web Voting," *San Jose Mercury News*, October 26, 2000.
[42] For example, see Tillett, L. Scott, "Will Internet Improve Voting?" Internet Week Online, November 17, 2000, http://www.internetweek,com/lead/lead111700.htm.

California has taken a more cautious approach.  Citing the security concerns of the Internet Voting Task Force report issued in January 2000, Governor Gray Davis vetoed a bill in September that would have authorized binding trials of Internet voting in state and local elections.  Instead, prior to the November 2000 election, four California counties – Contra Costa, Sacramento, San Mateo and San Diego – conducted non-binding tests of Internet voting from computers located at polling places.  According to VoteHere.net, the firm administering the trials in Sacramento and San Diego counties,[43] voters found the system easy to use, "8 out of 10 said they preferred Internet voting to the current system, and … 65 percent said they would vote from home if they thought the system was secure."[44]

As a result of the slow counts and other problems encountered with absentee ballots in the November 2000 election, some Internet voting advocates are now focusing on allowing absentee voters to use the Internet rather than the mails. This would be consistent with the conclusion reached by the Internet Voting Task Force that "it is technologically possible to utilize the Internet to develop an additional method of voting that would be at least as secure from vote-tampering as the current absentee ballot process in California."[45]  Improving the security and integrity of absentee voting seems a high priority for election reform,[46] which may create an opening for early tests of Internet voting by absentees.  Given that the percentage of California absentee ballots has grown from 6 percent in 1980 to 24.5 percent in 2000,[47] Internet voting would have the potential to grow rapidly once authorized. Oregon, where the November 2000 election was conducted entirely by mail, is also looking into the possibility of online voting.

---

[43] Safevote, Inc. and Election Systems and Software ran the Internet voting trials in Contra Costa and San Mateo counties, respectively.
[44] Schwartz, John, "E-Voting: Its Day Has Not Come Just Yet," *The New York Times*, November 27, 2000 <http://www.nytimes.com/2000/11/27/technology/27CHAD.html>.
[45] *Final Report of the California Internet Voting Task Force*, op. cit., p.1.
[46] Absentee voting has relatively little protection against fraud and other abuses. See, for example, Simpson, Glenn R. and Evan Perez, "'Brokers' Exploit Absentee Voters; Elderly Are Top Targets for Fraud," The Wall Street Journal, December 19, 2000.

**Internet Voting in Non-Government Elections**

Meanwhile, Internet voting has found new niches in the private and nonprofit sectors. Many publicly traded U.S. corporations, which are required to conduct annual shareholder elections for directors and on other proposals, now permit and encourage proxy voting over the Internet. The number of investors voting online has more than doubled each year for the past three years and in 2000 constituted about 15 percent of all voting shareholders.[48]

Other organizations such as credit unions, labor unions, professional societies, and university student governments are beginning to hold their elections online. Probably the largest such effort to date was the October 2000 direct election of five at-large members to the international governing board of the nonprofit Internet Corporation for Assigned Names and Numbers (ICANN). The Markle Foundation gave $500,000 to ICANN and other organizations to support the Internet vote, which was managed by election.com. Anyone at least 16 years old could register with ICANN by providing a permanent mailing address and e-mail address. ICANN then mailed an encrypted PIN to the individual, which functioned much like a digital signature to verify that the person was registered when he or she logged on to vote.

---

[47] Bustillo, Miguel, "Rise in Use of Absentee Ballot Alters Tactics as Election Day Nears," *Los Angeles Times*, November 3, 2000; Secretary of State, State of California, "Jones Officially Certifies California Election Results," December 15, 2000.
http://www.ss.ca.gov/executive/press_releases/2000/00-131.htm.
[48] Nathan, Sara, "More Investors Click To Cast Proxy Votes," *USA Today*, March 27, 2000.

Of the 76,000 individuals who registered as ICANN at-large members, 34,035 or nearly 45 percent voted during the 10-day voting period. Frank Fatone, Chief of Election Services for election.com, commented: "45% represents a significantly higher turnout than other private sector elections… We usually see 13-18%…turnout in elections of this type. Use of the Internet clearly had a positive impact on participation in the ICANN election."[49] However, some technical glitches occurred:

> During the first twelve hours of the 10-day voting period, some 2,800 of the 76,000+ At Large members encountered an error message when attempting to submit their votes. The difficulty was caused by the interaction of election.com's voting system with ICANN's encryption routine… The situation was identified and corrected within the first 12 hours of the voting period. ICANN members that were affected by the situation were notified immediately via e-mail, and were directed to log on and cast their vote. Of the 2,800 people who received an error on their first attempt, 2,685 returned to the site and successfully cast their votes.[50]

The ICANN election shows that Internet voting with digital signatures can work with large numbers of dispersed voters, but also that technical problems are likely to arise in the early implementations. These problems would have to be solved before online voting is used widely in binding government elections. As Zoe Baird, president of the Markle Foundation, said afterwards: "[The ICANN election was] far from perfect…It is now imperative that the data from this election experiment be thoroughly analyzed and available for public scrutiny so that the dialogue can continue and the system can be improved."[51]

---

[49] election.com, "ICANN and election.com Announce Results for First Worldwide Online Vote," October 10, 2000, http://www.election.com/us/pressroom/pr2000/1010.htm.
[50] Ibid.
[51] Markle Foundation, "ICANN Elections: An Important Moment for Internet Governance," October 11, 2000, http://www.markle.org/news/Release.200010111248.1872.html.

Managing Internet voting for corporations and non-government organizations represents an important near-term source of learning and revenue for electon.com and other firms such as Election Systems and Software, Safevote, Inc. and VoteHere.net. These firms expect to apply their experience to online government elections, and they would be well positioned to bid on support contracts for Internet petition signing as well.

**Online Security for E-Commerce and E-Government Applications**

Despite well-publicized failures of online retailers, Internet shopping continues to grow. A UCLA survey conducted in Spring 2000 found that more than half (51 percent) of U.S. Internet users have made purchases online.[52] The PPIC survey in October 2000 reported that 59 percent of California adults who use the Internet went online "to purchase goods or services."[53] For many young (and some older) adults, Internet shopping has become a familiar part of their daily lives.

As consumer online purchasing expands, e-commerce firms are seriously investing in identification and encryption to enhance security and generate customer trust. Shopping websites typically use registered passwords for identification and "secure socket layer" (SSL) encryption for transmitting credit card or other payment information.[54] Websites that offer high-value transactions such as securities purchases, mortgages, and insurance may add PKI digital signatures backed by third-party certification authorities to verify customers' identities. As a next step, online identification systems using biometric methods to recognize fingerprints, faces, or voices are under development and appear likely to find acceptance among consumers.[55]

---

[52] "The UCLA Internet Report: Surveying the Digital Future," UCLA Center for Communication Policy, November 2000, p. 10, http://www.ccp.ucla.edu.
[53] Public Policy Institute of California "PPIC Statewide Survey: Californians and Their Government--October 2000," op. cit., p. 28.
[54] Computer Science and Telecommunications Board, *The Internet's Coming of Age*, op. cit., p. 5/16.

Over the next few years, digital signatures and certification authorities developed for e-commerce will likely be used for such e-government applications as filing taxes, obtaining licenses or permits, and bidding for government procurement contracts, which still require written signatures. This change may require government approval of the certification authorities used in these transactions, which the office of the California Secretary of State has already initiated under its 1998 regulations.[56] Japan is also preparing regulations for ministerial approval of "certification services" under its recently passed digital signature law.[57]

Europe is well ahead of the United States in its use of smart cards for e-commerce and e-government applications. The European Commission is overseeing a formal plan to develop smart card requirements for a common "European Citizen Digital ID Document." According to one Commission report, this development

> …will promote European commerce and online payments. Moreover, it will be a very important step towards e-government in the European member states. Another benefit is enhanced data security. The qualified citizen's certificate enables strong authentication, encryption and digital signatures.[58]

Europeans have been more comfortable than Americans with government identity cards, and the European Citizen Digital ID Document represents both a modernization and harmonization of existing national paper ID documents into a common European Union digital format. No similar trend toward using smart cards for identification is apparent in the U.S., although credit card issuers continue to experiment with them.[59] It is quite possible that the U.S. credit card industry will replace existing magnetic-stripe cards with smart cards sometime within this decade, in large part to improve security for online transactions. However, the actual timing of such a move is difficult to predict.

---

[55] See, for example, Power, Carol, "Consumers Favor Fingerprint Scans in ID-Verification Tests," *American Banker*, December 22, 2000.
[56] Secretary of State, "California Digital Signature Regulations, op. cit.
[57] Government of Japan, "Law Concerning Electronic Signatures and Certification Services," enacted May 24, 2000. <http://www.miti.go.jp/english/special/E-Commerce/index.html>.
[58] Information Society Technologies Programme, "eEurope Smart Cards: Common Requirements," Brussels, European Commission, December 11, 2000, §7.1.1, http://europa.eu.int/ISPO/istka2/steeringmeetings.html.

---

[59] In September 1999, American Express launched "Blue," a smart card targeted to "technology-minded individuals." As of December 2000, it appears to have had only modest success. See "American Express Launches Blue," September 8, 1999, http://home3.americanexpress.com/corp/latestnews/blue.asp.

## 5. DISCUSSION, CONCLUSIONS AND RECOMMENDATIONS

To this observer, Internet petition signing does not yet seem ready for implementation in California or other states, but pressures for it seem likely to increase as more people use the Internet regularly to pursue their personal and professional interests, e-commerce, and interactions with government.

**Current Obstacles and Ameliorating Trends**

Security, access, and cost remain the principal obstacles to implementation of Internet petition signing. The security concerns associated with signing a petition on a remote computer are very real and appear difficult, but not impossible, to resolve satisfactorily. The continuing growth of e-commerce and new experiments with Internet voting will bring with them considerably more experience with digital signatures, biometrics, and other security approaches over the next few years. Given the commercial pressure to reduce risks and losses from large numbers of online transactions, identification and security methods will undoubtedly improve, and it seems highly likely that the commercial world will find workable solutions. Whether and when such solutions will be adequate to maintain public trust in remote signing of initiative petitions remains to be seen.

As costs decrease and a new, Internet-savvy generation reaches voting age, equity and access concerns will diminish but not disappear. Market and demographic forces alone will not bring all adults online. Consequently, any decision to permit Internet petition signing should include access arrangements for those who are not connected at home, school, or work. These arrangements would be consistent with the recommendations of the California Internet Voting

Task Force to provide Internet kiosks for registration or initiative signature gathering. They would also have obvious cost implications for government.

State and local government seems unlikely to pay for the needed security and access infrastructure solely for Internet petition signing. However, the growing interest in California and other states in using the Internet for government operations and services will go a long way toward building that infrastructure. In his State of the State address on January 8, 2001, Governor Davis officially launched a new state website -- <http://my.ca.gov> -- that provides a portal to e-government services such as registering vehicles, making state park campsite reservations, and checking the status of state income tax refunds. Hackers will surely test the privacy and security measures put in place for these e-government applications. As a result, it will be important to monitor, document, and analyze the ongoing security experience with e-government services, both to make these applications more secure and to inform any subsequent efforts to develop online voting or petition signing.

Election reforms in the aftermath of last November's problems may also have implications for petition signing. One such reform could be to update and maintain official voting lists online, with offline backup in the case of outage, intrusion, or other problems. Although initial voter registration would still require tangible proof of identity, such as a driver's license or social security card, subsequent changes could be processed online. Michigan's decision to link voter registration records to drivers' licenses, so that a DMV address change will automatically trigger a similar change on the voting rolls, also seems likely to spread to other states. Although such developments will not lead directly to Internet petition signing, they would provide much of the infrastructure needed for it.

The growth of remote Internet voting in the private and nonprofit sectors, along with more field trials in government elections, may further encourage other Internet applications in the political process such as petition signing. Despite the forecast by one well-respected consulting firm that "all states [will] have some form of Internet-based electronic voting by 2004,"[60] Internet voting must overcome many obstacles before it becomes widespread. Still, many voters say they favor online voting from home or work.[61] Moreover, absentee-voting reforms may include steps toward Internet voting. The issues surrounding Internet voting are closely intertwined with those for Internet petition signing, and future studies of or proposals for Internet voting should therefore consider the implications for initiative signature gathering on the Internet.

**Broader Impacts of the Internet on the Initiative Process**

Although the real effects of Internet signature gathering on the overall initiative process are as yet unknown, its proponents and opponents have focused on a few key points. Proponents have emphasized the Internet's potential to lower the cost and reduce the time required to qualify an initiative. Opponents usually stress the security and access concerns discussed above. Beyond these issues, however, lie more philosophical questions about how the Internet might influence initiatives and direct democracy generally.

One important question is whether the Internet could improve the quality of, as well as voters' actual use of, information about initiatives. Critics of the initiative process cite the scarcity and superficiality of information available to voters on television and radio.[62] In principle, the Internet is an ideal medium for presenting detailed information about specific initiatives and the groups supporting or opposing them. Internet websites can also link this information to relevant commentaries and other sources. Voters who seek information in greater depth than ballot pamphlets[63] and the mass media provide would be able to find it on the Internet.[64] As one example, California now requires all committees supporting and opposing ballot propositions that raise or spend $50,000 or more to file lists of contributors and contributed amounts electronically. This information is then made publicly available on the Secretary of State's website.[65]

---

[60] Gartner Group, " Gartner Says All States in the United States to Have Some Form of Internet-Based Electronic Voting by 2004," April 11, 2000, http://www.gartner.com/public/static/aboutgg/pressrel/pr041100b.html.
[61] "A poll by ABC News found that 61% of 18-34-year-olds would like to vote online." Chambers, op. cit.

A related question is whether and to what extent the Internet will encourage greater and more informed public participation in the initiative process. Initiative websites could include interactive message boards that stimulate public discussion and debate, as other websites now offer on nearly every conceivable topic. It is certainly true that website message boards often spiral down into banal chatter or diatribe; nevertheless, many examples of sustained, spirited discussions on serious topics also can be found. The Internet's capacity to allow substantial numbers of people interact over an extended period of time could counter another central criticism of initiatives: that they do not foster a structured, deliberative political process so essential to representative democracy.

---

[62] Broder, David, *Democracy Derailed*, New York: Harcourt, 2000; Frickey, Philip P., "Representative Government, Direct Democracy, and the Privatization of the Public Sphere," *Willamette Law Review*, 34, 421, 1998. See also Cronin, Thomas, *Direct Democracy: The Politics of Initiative, Referendum and Recall*, New York: Twentieth Century Fund, 1989.

[63] A California ballot pamphlet for the November 2000 election, with information about each initiative, was available online before the election at < http://vote2000.ss.ca.gov/VoterGuide>. The state also spent about $6.5 million to mail pamphlets to 10 million voter households. Harwood, op. cit.

[64] As of August 2000, 29 percent of California adults reported they went online "to visit the web sites of elected officials, political candidates, political parties, or political causes." Public Policy Institute of California, "PPIC Statewide Survey: Californians and Their Government--August 2000," http://www.ppic.org/publications/CalSurvey13/survey13.pdf.

[65] California Automated Lobbying and Campaign Contribution & Expenditure Search System (CAL-ACCESS), http://CAL-ACCESS.ss.ca.gov.

An interesting recent proposal would use the Internet for public discussion of initiatives during the drafting process so that the proposed language could be debated and modified before seeking ballot qualification.[66] This proposal would require major changes in the current legislation governing initiatives as a way of developing a forum "in which the mix of professional and public voices could create a deeply deliberative process of public law."[67] Of course, others will make precisely the opposite argument, contending that the Internet favors non-deliberative, emotional responses that only exacerbate the flaws of initiatives and other tools of direct democracy. In all likelihood, the Internet can and will be used in both ways simultaneously.

Perhaps the most significant question raised by Internet petition signing is whether its chief effect would be to worsen current problems surrounding the initiative process itself. Lowering the cost to qualify an individual initiative could inundate voters with ballot measures at every election and might, in fact, increase the total sum spent on initiatives. Along with sheer number of items to be voted on, the influence of money and organized interest groups could increase.[68]

Such concerns about intensifying the negative aspects of direct democracy, like the hopes for a positive Internet role in spurring informed public participation, are conjectural. We lack good data or systematic studies on these points[69] and simply do not understand the full implications of using the Internet for petition signing or voting.  The Internet can help level the political playing field among candidates and initiative proponents, but it could also exacerbate the influence of well-heeled contributors and organized interest groups.  It can inform and encourage participation among voters in ways other media cannot, but it could also stimulate and reward superficial, emotional responses.  It can be used for serious deliberation and debate on proposed initiatives among informed citizens, but it could also lead to an explosion of easy-to-qualify ballot measures with disastrous results for representative government.

---

[66] Worthington, Jay, "A Wider Hillside: Direct Democracy, Information Deficits and the Net," unpublished manuscript, 2000, pp. 27-28.
[67] Ibid., p. 28.
[68] As one example noted by a reviewer of an earlier draft of this paper, a well-financed group could pay individuals to place messages on initiative websites and thereby spin the discussion toward the group's point of view.
[69] Bimber, Bruce, "The Internet and Political Transformation: Populism, Community and Accelerated Pluralism," *Polity*, Vol. XXXL, No. 1, Fall 1998, pp. 133-160. For examples of speculative scenarios about the Internet and direct democracy, both positive and negative, see Corrado, Anthony and Charles M. Firestone, eds, *Elections in Cyberspace: Toward a New Era in American Politics*, Washington, D.C.: The Aspen Institute, 1996.

We can be fairly sure, however, that Internet signature gathering, like Internet voting, will have unintended consequences.  That prospect may be reason enough for many to oppose its early implementation in California, but it will not make the concept disappear.  Its proponents will likely gain strength as more young people who have grown up with the Internet reach voting age and see no reason why they should not engage in political activities online as they do in all other areas.

Internet petition signing seems an idea whose time is not yet ripe but is clearly ripening.  Its emergence on the political horizon should spur reformers of the initiative process to get on with their work before they are overtaken by events in cyberspace.

# REFERENCES

Bikson, Tora K. and Constantijn W. A. Panis, *Citizens, Computers and Connectivity*, Santa Monica, CA: RAND, MR-1109-MF, 1999.

Bimber, Bruce, "The Internet and Political Transformation: Populism, Community and Accelerated Pluralism," *Polity*, Vol. XXXL, No. 1, Fall 1998, pp. 133-160.

Broder, David, *Democracy Derailed*, New York: Harcourt, 2000.

Bustillo, Miguel, "Rise in Use of Absentee Ballot Alters Tactics as Election Day Nears," *Los Angeles Times*, November 3, 2000.

California Automated Lobbying and Campaign Contribution & Expenditure Search System (CAL-ACCESS), http://CAL-ACCESS.ss.ca.gov.

Caltech/MIT Voting Technology Project, "Voting: What Is and What Could Be," http://web.mit.edu/newsoffice/nr/2001/VTP_report_all.pdf.

Chambers, John, "Can Technology Fix Balloting Problems? Yes; Harness Strength of the Internet," *USA Today*, December 19, 2000.

Cooper, Audrey, "Legislator Proposes Online Voting for California," Associated Press, December 5, 2000.

Computer Science and Telecommunications Board, *The Internet's Coming of Age*, Washington, DC: National Academy Press, 2000, pp. 5/15-19.

________, *Trust in Cyberspace*, Washington, DC: National Academy Press, 1999, pp. 124-132.

Corrado, Anthony and Charles M. Firestone, eds, *Elections in Cyberspace: Toward a New Era in American Politics*, Washington, D.C.: The Aspen Institute, 1996.

Cortese, Amy, "The Software Revolution: The Internet Changes Everything," *Business Week*, December 4, 1995, http://www.businessweek.com/1995/49/b34531.htm.

Cronin, Thomas, *Direct Democracy: The Politics of Initiative, Referendum and Recall*, New York: Twentieth Century Fund, 1989.

Culp, Scott, "The Ten Immutable Laws of Security," October 2000, www.microisoft.com/technet/security/10imlaws.asp.

Davis, Donald, "Where There's A Web, There's A Way," *CardTechnology.com*, October 2000, http://www.cardtech.faulknergray.com/news.htm.

Department of Defense, "Voting Over the Internet: Pilot Project Assessment Report," June 2001, http://www.fvap.ncr.gov/voireport.pdf.

election.com, "Arizonans Register Overwhelming Support for Online Voting," March 12, 2000, http://votation.com/us/pressroom/pr2000/0312.htm.

__________, "ICANN and election.com Announce Results for First Worldwide Online Vote," October 10, 2000, http://www.election.com/us/pressroom/pr2000/1010.htm.

Frickey, Philip P., "Representative Government, Direct Democracy, and the Privatization of the Public Sphere," *Willamette Law Review*, 34, 421, 1998.

Gartner Group, " Gartner Says All States in the United States to Have Some Form of Internet-Based Electronic Voting by 2004," April 11, 2000, http://www.gartner.com/public/static/aboutgg/pressrel/pr041100b.html.

Government of Japan, "Law Concerning Electronic Signatures and Certification Services," May 24, 2000, http://www.miti.go.jp/english/special/E-Commerce/index.html.

Harwood, John, "Fixing the Electoral System: Lessons From States Hold Hope for Reform," *The Wall Street Journal*, December 22, 2000.

"ICANN and election.com Announce Results for First Worldwide Online Vote," October 10, 2000, http://www.election.com/us/pressroom/pr2000/1010.htm.

Information Security Committee, "Digital Signature Guidelines," American Bar Association, Section of Science and Technology, Electronic Commerce and Information Technology Division, 1996, http://www.abanet.org/scitech/ec/isc/digital_signature.html.

Information Society Technologies Programme, "eEurope Smart Cards: Common Requirements," Brussels, European Commission, December 11, 2000, <http://europa.eu.int/ISPO/istka2/steeringmeetings.html>.

"Internet Access Tops 56 Percent in U.S.," Nielsen/Net Ratings, December 15, 2000 http://209.249.142.22/press_releases/PDF/pr_001215.pdf.

Internet Policy Institute, "Report of the National Workshop on Internet Voting," March 6, 2001, http://www.internetpolicy.org/research/e_voting_report.pdf.

Jesdanun, Anick, "Resistance Continues for Web Voting," *San Jose Mercury News*, October 26, 2000.

Markle Foundation, "ICANN Elections: An Important Moment for Internet Governance," October 11, 2000, http://www.markle.org/news/Release.200010111248.1872.html.

Matthews, William, "Election Day Winner: Online Voting," Federal Computer Week, Nov. 10, 2000, http://www.fcw.com/fcw/articles/2000/1106/web-elect-11-10-00.asp.

Mercuri, Rebecca, "Electronic Voting," http://www.notablesoftware.com/evote.html.

Nathan, Sara, "More Investors Click To Cast Proxy Votes," *USA Today*, March 27, 2000.

National Telecommunications and Information Administration, *Falling Through the Net: Toward Digital Inclusion*, U.S. Department of Commerce, 2000, http://search.ntia.doc.gov/pdf/fttn00.pdfl.

Power, Carol, "Consumers Favor Fingerprint Scans in ID-Verification Tests," *American Banker*, December 22, 2000.

Public Policy Institute of California, "California's Digital Divide," November 2000, http://www.ppic.org/facts/digital.nov00.pdf.

Rubin, Avi, "Security Considerations for Remote Electronic Voting over the Internet," 2000. http://avirubin.com/e-voting.security.html.

Schneier, Bruce, "Why Digital Signatures Are Not Signatures," *CRYPTO-GRAM*, November 15, 2000, http://www.counterpane.com/crypto-gram-0011.html.

Schwartz, John, "E-Voting: Its Day Has Not Come Just Yet," *The New York Times*, November 27, 2000, http://www.nytimes.com/2000/11/27/technology/27CHAD.html.

Secretary of State, State of California, *Final Report of the California Internet Voting Task Force*, January 18, 2000, http://www.ss.ca.gov/executive/ivote/.

\_\_\_\_\_\_\_\_\_\_\_\_\_\_, "California Digital Signature Regulations, June 12, 1998, www.ss.ca.gov/digsig/regulations.htm.

\_\_\_\_\_\_\_\_\_\_\_\_\_\_, California Automated Lobbying and Campaign Contribution & Expenditure Search System (CAL-ACCESS), http://CAL-ACCESS.ss.ca.gov.

"Jones Officially Certifies California Election Results," December 15, 2000, http://www.ss.ca.gov/executive/press_releases/2000/00-131.htm.

Shamos, Michael Ian, "Electronic Voting: Evaluating the Threat," 1993, http://www.cpsr.org/conferences/cfp93/shamos.html.

Simmons, Charlene, "California's Initiative Process: A Primer," California Research Bureau, California State Library, CRB-97-006, May, 1997.

Simpson, Glenn R. and Evan Perez, "'Brokers' Exploit Absentee Voters; Elderly Are Top Targets for Fraud," *The Wall Street Journal*, December 19, 2000.

Smart Initiatives Initiative, 2000, http://www.smartinitiatives.org/English/civiset.html.

Stellin, Susan, "ICANN Holds Public Election," *The New York Times*, October 12, 2000.

Strassman, Marc, "After Florida, What?" Smart Initiatives Online Newsletter, November 12, 2000.

\_\_\_\_\_\_\_\_\_\_\_\_ "Fuzzy Math for Smart Initiatives," Smart Initiatives Online Newsletter, December 14, 2000.

Tillett, L. Scott, "Will Internet Improve Voting?" Internet Week Online, November 17, 2000, http://www.internetweek,com/lead/lead111700.htm.

"The UCLA Internet Report: Surveying the Digital Future," UCLA Center for Communication Policy, November 2000, http://www.ccp.ucla.edu.

Voting Integrity Project, "Hurtling Toward Cyber-Elections," 1999, http://www.voting-integrity.org/projects/votingtechnology/internetvoting/ivp_3_hurtling.htm.

Worthington, Jay, "A Wider Hillside: Direct Democracy, Information Deficits and the Net," unpublished manuscript, 2000.